\documentclass{article}
\usepackage{amsmath}
\usepackage{amssymb}
\usepackage{mathrsfs}
\usepackage{graphicx}
\usepackage{pb-diagram,pb-xy}
\usepackage{lamsarrow}
\usepackage{pb-lams}
\def\beq{\begin{equation}}
\def\eeq{\end{equation}}
\newtheorem{note}{Note}
\newtheorem{defn}{Definition}
\newtheorem{rem}{Remark}
\newtheorem{lem}{Lemma}
\newtheorem{prop}{Proposition}
\newtheorem{thm}{Theorem}
\newtheorem{cor}{Corollary}

\newcommand{\aij}[3]{{#1}^{#2}_{{\phantom #2}#3}}

\newcommand{\bn}{{\mbox{\boldmath $\omega$}}}
\newcommand{\ba}{{\bf a}}
\newcommand{\bp}{{\bf p}}
\newcommand{\bq}{{\bf q}}
\newcommand{\bx}{{\bf x}}
\newcommand{\by}{{\bf y}}

\newcommand{\cav}{{\mathcal V}}

\newcommand{\hf}{{\hat\phi}}

\newcommand{\BR}{\mathbb{R}}
\newcommand{\pf}{{\bf Proof}. }

\newcommand\qed{\begin{flushright}$\square$\end{flushright}}
\begin{document}
%\begin{frontmatter}
\title{The Theory of Kairons}
\author{Arkadiusz Jadczyk\footnote{E-mail address: arkadiusz.jadczyk@cict.fr}\smallskip \\ \emph{Center CAIROS, }\\ \emph{Institut de Math\'{e}matiques de
    Toulouse,}\\ \emph{Universit\'{e} Paul Sabatier,}\\ \emph{31062 TOULOUSE CEDEX  9, France }}
%\date{\today}
\maketitle
\begin{abstract}
In relativistic quantum mechanics wave functions of particles satisfy field equations that have initial data on a space--like
hypersurface. We propose a dual field theory of ``wavicles'' that have their initial data on a time--like worldline. Propagation
of such fields is superluminal, even though the Hilbert space of the solutions carries a unitary representation of the Poincar\'{e}
group of mass zero. We call the objects described by these field equations ``Kairons''. The paper builds the field equations
in a general relativistic framework, allowing for a torsion. Kairon fields are section of a vector bundle over space-time. The bundle has infinite--dimensional fibres.
\end{abstract}
%\begin{keyword}

%\PACS
%\end{keyword}
%\end{frontmatter}
\section{Introduction}
Einstein's Special Theory of Relativity united space and time into one space--time continuum. Julian Barbour in his book "The End of Time" \cite{barbour}
proposed to go even further by building a whole philosophy around the idea of timelessness. Yet the fact is that what we human beings perceive, what counts, is not timelessness and not even the continuous, linear clock time. What counts for human beings are "events", irreversible discontinuities in time. This was the view of Ilya Prigogine \cite{prigogine} who has stressed the need for second time, "time of becoming" in contrast to the ordinary time, time of "being". But how to implement this idea in physics?

Prigogine suggested that irreversibility is somehow implanted into the fundamental laws of microphysics. Yet it seems that we are still lacking the relevant mathematical structures, structures that go beyond "master equations", structures that apply to
the very ways of how we talk about the physical reality.

The present paper is an attempt at constructing, theoretically, such a new structure. This effort resulted from observing a natural duality between space and time, and by exploiting this duality. Because time is one--dimensional, while space is three--dimensional, this duality is not of a kind that can be immediately seen at the classical level. Here time serves as a parameter for the dynamics, space is an arena where the dynamics is taking place. And yet this duality becomes apparent when we go to the quantum description level.\footnote{Cf. {\em ``Relativistic Quantum Events''\,}, Ref. \cite{blaja96}.}

Mathematical investigations of the structure of quantum theories have led, starting from Birkhoff and von Neumann, to the concept of ``quantum logic'', a noncommutative generalization of the classical logic of Aristotle and Boole. There are different ways of constructing examples of non-Boolean logics, one of them being via the concept of ``orthogonality''.

Normally, when discussing orthogonality, we have in mind orthogonality of vectors and vector subspaces in a Hilbert space. But it does not have to be so. We can, for instance, consider events in Minkowski space $M$ and call two events, $x$ and $y$ ``orthogonal", $x\bot y,$ if they are can not be connected by a time--like interval. This leads to a non--Boolean logic $(M,\bot)$ that has the essential properties of a quantum logic - it is a complete orthomodular lattice.

Since the orthogonality relation is invariant under the Poincar\`{e} group, we get a covariant logic and we can look for covariant representations of this logic, where the Poincar\'{e} group operations will be represented by unitary operators. The simplest covariant
representation of this ``causal logic of the Minkowski space'' can be constructed from the solutions of a massless free Dirac equation. But we can also consider a dual orthogonality relation $x\top y$ that holds if an only if $x\neq y$ and $x$ is time--like or light--like with respect to $y.$ This relation also leads to a non-Boolean logic but $(M,\top)$, satisfies a somewhat weaker axioms than $(M,\bot).$ It is an ortho--modular partially ordered set, but not a lattice \cite{ceja77}.

The term ``Kairons'' has been chosen for naming the wavicles giving rise to this reversed space--time logic in reference to one of the two important Greek gods of time. The standard, linear and continuous time is associated with the name of the ``dancer'' time -- Chronos, while the
god of the discontinuous time, the ``jumper'', is called Kairos \footnote{More on this subject in the forthcoming paper {\em ``Some aspects of contemporary Kairicity\,}'' by P.~Ang\`{e}s and the present author} The natural question that appears is: what kind of a field equations lead to covariant representations
of $(M,\top)$?

In order to answer this question it is necessary to realize that the key element is the ``probability current''. In the case of the Dirac equation (massive or massless) the probability current is given by the sesquilinear form $j^\mu = {\bar \Psi}\gamma^\mu\Psi$ that is ``conserved": $\partial_\mu j^\mu=0.$ Such a representation of the probability current, that is standard in physics books, is somewhat misleading. In fact, the probabilistic representation works also for a massless Dirac equation that is conformally invariant. In such a case what we get naturally from the geometry is not
a vector--valued current, but a $3$--form $j$ that is closed $dj=0.$ That means that (for solutions that have compact support on space-like hypersurfaces) the integral of $j$ over a space--like hypersurface does not depend on the choice of this hypersurface; the physical interpretation of this fact reads: ``the particle moves along a time--like worldline and will be detected with certainty by any instant measurement determining its presence.''

If we want to have a dual picture, where the roles of "space--like'' and ``time--like''are reversed, we need not a particle but a ``wavicle'', an object located on a hypersurface that is intersected by any ``observer's'' time--like worldline. While a particle is a {\em singularity in space\,}, a wavicle must be a {\em singularity in time\,}. For this we need a current that
is a closed $1$--form, not a closed $3$--form as it is for particles. It is clear that it is rather impossible to deduce field equations leading to such a current from an action principle. A direct approach is necessary. This is the approach taken in the present paper.

As a template we take
a pseudo-Euclidean space $E^{(1,m)}$ of signature $(1,m).$ The main object is the positive light cone $C_0^{+*}$ and its projective image isomorphic to an $(m-1)$--dimensional sphere $S.$ In this paper we work over the field of real numbers $\BR.$

Sec 2 is devoted to the recall of algebraic constructions that are taking place in one fibre of a bundle. Bundles are discussed in Sec. 3.

In Sec. 2.1. we introduce the space of frames $F(V)$
of a vector space $V$ of dimension $m,$  its subset $P$ which is a reduction of $F(V)$ to a subgroup $G\subset GL(m),$ and discuss the space $P\times_RQ$ of geometric quantities of type $R,$ where $R$ is a left action of a $G$ on a space $Q.$

In Se. 2.2 we represent geometric quantities of type $R$ as equivariant functions on $P.$ All this is standard and is being recalled in order to fix our notation used in the sequel.

In Sec. 2.3 we
take $V$ to be $(m+1)$ dimensional and specify $G$ as $SO_0(1,m)$ -- the generalized Lorentz group. The important objects are the positive light cones
$C_0^{+*}\subset E^{(1,m)}$ and $C0^{+*}\subset V.$ We introduce the invariant measure on $C_0^{+*}$ and the action $\Lambda\rightarrow \rho_\Lambda$ of $SO_0(1,m)$ on the projective light cone isomorphic to the sphere $S\subset\BR^{m-1}.$ This action dtermines a family of cocycles $\gamma_r,\, r\in \BR$ on $SO_0(1,m)\times S$ that are used later for the construction of a family of infinite dimensional representations $R_r$ of $SO_0(1,m).$ We also
analyze the $SO_0(1,m)$--noninvariance of the canonical $SO(m)$ invariant measure on $S.$ Representations $R_r$ on $C^\infty (S)$ and, more generally, on spaces of vector valued differential forms on $S$ are discussed in Sec. 2.4.

In Sec. 2.5  We construct spaces ${\mathcal Y}_r^p$ of equivariant $p$--forms on $P$ with values in a vector space $W$ and discuss several important examples of elements of these spaces. An invariant integration over the sphere $S$ is introduced in Theorem 1. This integration is used later on for the construction of the conserved ($dj=0$) current.

In Sec 2.6 we interpret ${\mathcal Y}_r^0(S;\BR)$ as spaces of homogeneous functions of degree $-r$ on the positive light cone $C^{+*}\subset V.$ In Sec. 3 we move from algebra to geometry by introducing an $(m+1)$--dimensional manifold $M$ endowed with an $SO_0(1,m)$ structure and a compatible principal connection, possibly with a non--zero torsion. In fact, we slightly generalize our scheme allowing also for a degenerate space--time metric.
After recalling some important properties of the exterior covariant covariant derivative in Proposition 3, we construct, in Sec. 3.2, the Kairon bundle ${\mathcal Y}[P].$ In Proposition 4 several equivalent ways of interpreting cross sections of this bundle are given. In Sec 3.3 we comment on the generalized torsion, and in Sec. 3.4 we introduce the field equations (Eq. (31) and prove (Proposition 5) that these field equations lead to the conservation (Eq. (32)) of the current $j_{\Psi_1,\Psi2}$ defined by integration over the sphere of a form that is bilinear in the solutions of the field equations.

In Sec. 4 we specialize to the case of $M$ being the flat Minkowski space $E^{(1,m)}.$ The Kairon field is then described by a (real valued) function $\Psi(x,\bn),$ where $x$ is a point in $M$ and $\bn$ is a ``space direction'' at $x.$ The field equation are now reduced to a simple
form given in Eq. (33). In Proposition 6 the initial value problem is solved, where it is shown that each solution $\Psi(x,\bn)$ is uniquely determined by its values $\Psi(\gamma(s),\bn)$ on an arbitrary time--like worldline $\gamma(s).$ The field propagates along isotropic hyperplanes, therefore
its propagation is superluminal\footnote{For a discussion of superluminal solutions of massless field equations cf. also \cite{rodvaz} }.

In Sec. 4.2 the conserved current is used for the construction of
the (real) Hilbert space of solutions. Poincar\'{e} invariance is studied in Sec. 4.2, where it is shown that this Hilbert space carries a natural unitary representation of the Poincar\'{e} group.

This paper will be purely mathematical. A possible physical interpretation of the results as well as a generalization to the case of Spinning Kairons, using Clifford algebraic techniques,  will be given in a forthcoming paper.
\section{Algebraic Preliminaries}
We will be working in the smooth category, so that all manifolds, maps, and actions will be assumed to be smooth.
All vector spaces in this paper will be over the field of real numbers
$\BR.$ We denote by $\BR^+$ the multiplicative semigroup of strictly positive real numbers. If $M$ is a manifold, we will denote by $C^\infty(M)$ the space of smooth $\BR$--valued functions on
$M$ and by $\bigwedge^p(M)$ the $C^\infty(M)$--module of differential $p$--forms on $M.$  If $W$ is a vector
space, we will denote by $\bigwedge^p(M;W)$ the $C^\infty(M)$--module of $W$--valued $p$--forms on $M.$ If $f$
is a map between manifolds $f:M\rightarrow N$ then $f^*:\bigwedge^p(N;W)\rightarrow \bigwedge^p(M;W)$ is the
pullback map.

When dealing with fiber bundles, there are a number of constructions that are taking place in each fiber separately, and usually deal with algebra only. In order for this paper to be as self--contained as possible we provide these
algebraic preliminaries and separate them from the rest of the text in the following subsections.
\subsection{Geometric quantities of type $R$}\label{sec:gq}Let $V$ be a vector space of dimension $m.$ We denote by $F (V)$ the space
of linear frames of $V.$ The general linear group $GL(m)$ acts transitively and freely on $F (V)$ from the right:
\[GL(m)\ni A: e=(e_i)\mapsto eA=(e_j\,\aij{A}{j}{i}).\]
If $G$ is a subgroup of $GL(m),$ then an orbit $P$ of $G$ in $F (V)$ is called a $G$--structure on $V.$

Let $P$ be a $G$--structure on $V,$ and let $R$ be a left action of $G$ on a manifold $Q.$ On the direct product manifold $P\times Q$ one can
then introduce the equivalence relation: \beq (e,p)\sim (e',p')\quad \Leftrightarrow\quad \exists\, A\in G\,\,
\mbox{such that}\,\,((e'=eA)\,\wedge\, (p=R(A)p')).\eeq Denoting by $e\cdot p$ the equivalence class of $(e,p)\in P\times Q,$
we thus have \beq eA\cdot p=e\cdot R(A)p,\quad \forall\, A\in G.\label{eq:classes}\eeq The set of all such equivalence
classes is denoted by $P\times_R Q,$ and its elements are called {\it geometric quantities of type $R$\,}
\cite[Ch. II.6]{sternberg}.
\vskip5pt\noindent{\bf Example:} For instance, let $GL_+(m)$ denote the subgroup of $GL(m)$ consisting of all $m\times m$ matrices of positive determinant.
Then a $GL_+(m)$ structure is called an orientation of $V.$ Denoting by $\BR_+$ the
multiplicative group of positive real numbers let $\rho_w$  be the action of $GL_+(m)$ on
$\BR$ defined by \beq \rho_w: GL_+(m)\ni A\mapsto \det(A)^{w}\in
\BR.\eeq Let $F_+\subset F(V)$ be a fixed orientation. Given a real number $w,$ let $V^w$ denote the space $F_+\times \rho_w\, \BR,$
associated to $F_+$ via the representation $\rho_w.$ Elements of $V^w$ are called {\em densities of weight $w$\,}.\footnote{Our definition of the weight differs by the sign
from the one used  by Schouten \cite[Ch. II.8]{schouten}.} Every
oriented frame $e\in F_+$ defines an oriented $m$--vector
$e_1\wedge\ldots\wedge e_m.$ Let $\Lambda^m_+$ denote the set of all such  $m$--vectors. Then $\Lambda^m_+\simeq\BR_+$ and $\BR_+$ acts freely and transitively on $\Lambda^m_+$ by multiplication. It follows that $V^w$ can be also
considered as the space associated to $\Lambda^m_+$ via the
action $\BR_+\ni x: y\mapsto x^{w}y,\, y\in\BR_+.$ \vskip5pt\noindent
Any algebraic or geometrical structure of $Q$ that is invariant under the action $R$
of $G$ can be transported from $Q$ to $P\times_R Q.$ In particular, if $Q$ is a vector space $W$, and if $R$ is a
linear representation of $G$ on $W,$ then $P\times_R W$ inherits from $W$ the vector space structure of the same
dimension as $W.$ If $\{E_i\}$ is a basis in $W,$ then, for every frame $e\in P,$ the vectors $e_i=e\cdot E_i$
form a basis in $P\times_R W.$

Let us recall that if we choose $G\subset GL(m),$ $W=\BR^m,$ and if $R$ is the natural action of $G$ on
$\BR^m:$
\[(R(A)x)^i=\aij{A}{i}{j}\,x^j,\] then $F(V)\times_R W$
is naturally isomorphic to $V.$ If we choose $W=\BR^{m*},$ and if $R^\prime$ is the natural representation of
$G$ on $\BR^{m*}$: \[ (R^\prime (A)y)_j=y_i\, \aij{(A^{-1})}{i}{j},\] then $F(V)\times_{R^\prime } W$ is
naturally isomorphic to $V^*$ -- the dual of $V.$
\subsection{Geometric quantities as equivariant functions}\label{sec:gqef}
Let $G$ be a subgroup of $GL(m)$ and let $P\subset F(V)$ be a $G$--structure. Let, as it was discussed above, $R$ be the
right action of $G$ on $Q.$ A function $\Phi:P\rightarrow
 Q,$ $P\ni e\mapsto \Phi[e]\in Q,$
is said to be {\em equivariant of type $R$\,} if \beq \Phi[eA]=R(A^{-1})\,\Phi[e].\label{eq:equivar}\eeq There is a one--to--one
correspondence $\Phi\mapsto \tilde{\Phi}\in P\times_R Q$ between equivariant functions on $P$ of type $R$ and
geometric quantities of type $R,$ that is elements of $P\times_R Q.$ If $\Phi:P\rightarrow Q$ is equivariant of
type $R$ then, as it can be easily seen, the class $P\times_R Q \ni \tilde{\Phi}=e\cdot \Phi[e]$, in fact,  does not depend on $e\in P.$ Conversely, if
$\tilde{\Phi}$ is in $P\times_R Q,$ then for each $e\in P$ there exists a unique $\Phi[e]\in Q$ such that
$\tilde{\Phi}=e\cdot \Phi [e].$ The $Q$--valued function $e\mapsto {\tilde\Phi}[e]$ is then, by the construction, equivariant of type
$R.$

In applications, in order to avoid cumbersome notation, it is sometimes convenient to suppress the notational
difference between geometric quantities interpreted as elements of $P\times_R Q$ or as equivariant functions on
$P.$ The exact meaning should in such a case be deduced from the context.
\subsection{The invariant measure on the light-cone}\label{sec:a1lc}
We will denote by $E^{(1,m)}$ (resp. $E^{(1,m)*}$) the space $\BR^{(m+1)}$ (resp. $\BR^{(m+1)*}$) endowed with the quadratic form $q$
\beq
q(p^0,p^1,\dots\, ,p^m)= (p^0)^2-(p^1)^2-\ldots\, -(p^m)^2.\eeq
We will use the same symbol $q$ for the induced dual quadratic form
\beq q(p_0,\bp)= p_0^2-p_1^2-\ldots\, -p_m^2, \label{eq:qs}\eeq
the meaning will be clear from the context. The form $q$ is invariant under the natural action
of the group $SO_0(1,m)$ - the connected component of the identity of the full invariance group $O(1,m)\subset GL(m+1)$ of $q.$
\begin{note} For brevity, in the following, $G$ will stand for $SO_0(1,m)$ and $S$ will stand for the unit sphere
$S^{m-1}\subset \BR^{m*}.$  From now on we will denote by $V$ a fixed $m+1$--dimensional vector space equipped with a $SO_0(m,1)$
structure $P$. \
\end{note}

Since the quadratic form $q$ is $G$--invariant, it
induces a quadratic form, which we will denote by $Q,$ on $V$ and on $V^*.$ We denote by $\langle \cdot ,\cdot\rangle$ the
associated symmetric bilinear form on $V$ and on $V^*$ of signature $(1,m).$ All frames $e\in P$ are then orthonormal with
respect to $\langle \cdot ,\cdot\rangle :$ $e\in P \Longrightarrow \langle e_\alpha ,e_\beta\rangle
=\eta_{\alpha\beta}=\mathrm{diag}(1,-1,\ldots,-1).$

Throughout the paper the Greek indices $\alpha,\beta,\mu,\nu,$ etc. will run through $0,\ldots\,,m,$ while, unless explicitly
specified otherwise, the Latin indices $i,j,k$ etc. will run through $1,\ldots\,,m.$ Bold symbols $\bp,\bq,$
etc. will be used for vectors in $\BR^m$ and in $\BR^{m*},$ while the symbol $\bn=(\omega_1,\ldots ,\omega_m)$ will be
reserved for vectors in $\BR^{m*}$ of unit norm. We will use the symbol $\omega$ to denote isotropic vectors of the
form $\omega=(1,\bn),$ so that $\omega_0=1,$ $\sum_{i=1}^m (\omega_i)^2=1.$

A typical basis $e\in P$ will be also denoted as $e_\alpha,$ the dual basis as $e^\alpha,$ and a typical vector
$p$ in $V^*$ will be decomposed with respect to such a basis as $p=p_\alpha e^\alpha.$ Most of our constructions
will take place in $V^*.$

If $\Lambda$ is in $G$ ($=SO_0(1,m)),$ and if $\bn\in \BR^{m*}$ is a unit vector, then $\omega\Lambda$ will denote the vector in $E^{(1,m)*}$
with components $(\omega\Lambda)_\alpha=\omega_\beta \aij{\Lambda}{\beta}{\alpha},$ that is
$(\omega\Lambda)_0=\aij{\Lambda}{0}{0}+\omega_i\aij{\Lambda}{i}{0},$ $(\omega\Lambda)_i=\aij{\Lambda}{0}{i}+\omega_j\aij{\Lambda}{j}{i}.$

Let $C_0^{+*}$ be the positive isotropic cone in $E^{(1,m)*}$:
\[C_0^{+*}=\{(p_0,p_1,\ldots\, ,p_m):\, q(p_0,\bp)=0,\,
p_0>0\}.\] $C_0^{+*}$ is naturally diffeomorphic to $\BR^{m*}\setminus \{0\},$ since it is uniquely parametrized
by the non--zero vectors $\bp=(p_1,\ldots\, ,p_m)\in\BR^{m*}.$ It is well known that the $m$-form \beq
\mu_0(\bp)=\frac{dp_1\wedge\ldots\,\wedge dp_m}{|\bp |}\label{eq:mu0}\eeq is invariant with respect to the action of
$G$ on $C_0^{+*}$ induced by its natural action on $E^{(m,1)}.$ To see that this is the case, notice first that,
owing to the transitivity of the $G$ action on $C_0^{+*},$ the invariant $m$--form $\mu_0$, if it exists, is
unique up to a scale. To fix the scale, with $q$ given by Eq. (\ref{eq:qs}), we impose the condition: \beq dq\wedge\mu_0(\bp) = 2dp_0\wedge\ldots\wedge dp_m\label{eq:muq}\eeq at the points of $C_0^{+*},$
 where we notice that $dp_0\wedge\ldots\wedge dp_m$ is naturally $G$--invariant. Then a simple
calculation shows that $\mu_0$ defined in (\ref{eq:mu0}) indeed satisfies (\ref{eq:muq}).

Owing to its invariance under the action of $G,$ the form $\mu_0$ defines an $m$--form $\mu$ on the positive
isotropic cone (with respect to the quadratic form $Q$) $C^{+*}\subset V^*.$ Explicitly, given a frame $P\ni
e=\{e_\alpha\}$ we have: \beq
\mu(p;\,\xi^{(1)},\ldots\,,\xi^{(m)})=\frac{\epsilon^{i_1\ldots\,i_m}\xi^{(1)}_{i_i}\ldots\,\xi^{(m)}_{i_m}}{p_0},\label{eq:mu}\eeq
where $p\in C^{+*},$ $p=\sum_0^m p_\alpha e^\alpha,$ $\xi^{(i)}\in T_p C^{+*}$ are tangent vectors to $C^{+*}$ at
$p,$ considered as vectors in the vector space $V^*,$ with coordinates $\xi^{(i)}_{\alpha},$
$\xi^{(i)}=\xi^{(i)}_0e^0+\sum_{j=1}^m \xi^{(1)}_j e^j,$ and $\epsilon^{i_1\ldots\,i_m}$ is the fully
antisymmetric Kronecker tensor, with $i_1,\ldots\,,i_m\in \{1,\ldots\, m\}.$ The invariance of the measure
$\mu_0$ is reflected by the fact that the value of the right hand side of Eq. (\ref{eq:mu}) does not depend on
the choice of a frame $e\in P.$

The linear action of the group $G$ restricts to the action on the cone $C_0^{+*},$ and thus induces an action on the projective cone that is
isomorphic to the sphere $S\approx\{p\in C_0^{+*}:\, p_0=1\}.$  We will denote this action by $G\ni \Lambda\mapsto
\varrho_\Lambda:S\rightarrow S\subset \BR^{m*}.$ Explicitly:
\beq\varrho_\Lambda(\bn)_i=\frac{(\omega\Lambda^{-1})_i}{(\omega\Lambda^{-1})_0}.\label{eq:rholambda}\eeq
\begin{defn}
A function $\gamma(\Lambda,\bn)$ on $G\times S,$ with values in $\BR,$ satisfying:
\begin{enumerate}
\item[{\rm (i)}]$\gamma(I,\bn)=1,\,\forall\, \bn\in S,$
\item[{\rm (ii)}]$\gamma(\Lambda_1\Lambda_2,\bn)=\gamma(\Lambda_1,\varrho_{\Lambda_2}(\bn))\gamma(\Lambda_2,\bn),\,\forall\,
(\Lambda_1,\Lambda_2\in G,\, \bn\in S)\label{def:coc}$
\end{enumerate}
is called a cocycle.
\end{defn}
\begin{rem} By putting $\Lambda_2=\Lambda$ and $\Lambda_1=\Lambda^{-1}$ it follows from \rm{(i)} and \rm{(ii)}
above that for a cocycle $\gamma$ and all $\Lambda\in G,\, \bn\in S$ the following formula holds: \beq
\gamma(\Lambda^{-1},\varrho_\Lambda(\bn ))=\gamma(\Lambda,\bn)^{-1}.\label{eq:cocinv}\eeq
\end{rem}
\begin{lem}
For every real number $r\in \BR,$ the function $\gamma_r: G\times S\rightarrow \BR^+$ given by the formula
\beq\gamma_r(\Lambda,\bn)=((\omega\Lambda^{-1})_0)^{r}\label{eq:gammar}\eeq is a cocycle, that is $\gamma_r$ satisfies
conditions $(i)$ and $(ii)$ in the Definition \ref{def:coc} above. Moreover, for any $r,s\in\BR$ we have
\beq\gamma_r\,\gamma_s=\gamma_{r+s}.\label{eq:gadd}\eeq\label{lem:gadd}\end{lem} \pf The proof follows by
a straightforward calculation, along the lines given in Ref. \cite[Lemma 4]{jad06c}.\qed\vskip10pt Let, for each
$e\in P, $ $\Phi_e$ be the map $\Phi_e:E^{(1,m)*}\rightarrow V^*$ given by: \beq \Phi_e(x)=
x_\alpha\,e^\alpha,\quad x\in E^{(1,m)*}.\eeq We will denote by $\phi_e$ the restriction of $\Phi_e$ to the
isotropic cone $C_0^{+*}\subset E^{(1,m)*}:$ \beq \phi_e (\bp )= |\bp|e^0+p_i\,e^i .\label{eq:phie}\eeq Finally,
we denote by ${\hat \phi}_e$ the restriction of $\phi_e$ to $S\subset \BR^{m*}:$ \beq {\hat \phi}_e:\,
S\ni\bn\mapsto e^0+\omega_i\,e^i\in V^*,\quad \bn^2=1.\eeq \beq
\begin{diagram}
\node{S}\arrow{s,t,J}{}\arrow{se,t}{\hf_e}\\
\node{C_0^{+*}}\arrow{e,l}{\phi_e}\arrow{s,t,J}{}\node{C^{+*}}\arrow{s,,J}\\
\node{E^{(1,m)*}}\arrow{e,b}{\Phi_e}\node{V^*}
\end{diagram}\label{eq:diag}\eeq
Notice that while the image $\phi_e(C_0^{+*})$ is the positive isotropic cone $C^{+*}$ in $V^*$ that is
independent of $e\in P,$ the image $\hf_e(S)\subset C^{+*}$ varies with $e.$
\begin{lem} For each $e\in P$ let $\sigma[e]={\hat \phi}_e^*(\mu)$ be the $(m-1)$--form
on $S$ defined by \beq
\begin{array}{rcl}\sigma[e](\bn;\, \zeta^{(1)},\ldots\,,\zeta^{(m-1)})&=&{\hat \phi}_e^*(\mu)(\bn;\zeta^{(1)},\ldots\,,\zeta^{(m-1)})\\
&=&\mu(\hf_e(\bn);d\hf_e(\zeta^{(1)}),\ldots\,,d\hf_e(\zeta^{(m-1)})),
\end{array}\label{eq:sigma}\eeq
where $\zeta^{(1)},\ldots\,,\zeta^{(m-1)}\in \BR^{m*}$ are vectors tangent to $S$ at $\bn.$ Then $\sigma[e]$ is,
in fact, independent of $e\in P,$ and is the standard, $SO(m)$ invariant volume form $\sigma_0$ on $S.$  For
$\Lambda\in G$ we have: \beq
(\varrho_\Lambda^*\sigma_0)(\bn)=\gamma_{1-m}(\Lambda,\bn)\,\sigma_0(\bn),\label{eq:rn}\eeq where
$\varrho_\Lambda^*\sigma_0$ is the pullback of $\sigma_0$ by $\varrho_\Lambda.$ \label{lem:sigma}\end{lem}
\pf It follows directly from the definition of ${\hat \phi}_e$ that $d\hf_e(\zeta)=\zeta_i\,e^i,$
and $(\hf_e(\bn))_0=\langle e_0,\hf_e(\bn)\rangle =1.$ Therefore, applying (\ref{eq:mu}), we get \beq
\sigma[e](\bn;\,\zeta^{(1)},\ldots\,,\zeta^{(m-1)})=\epsilon^{i_1\ldots\,i_m}\omega_{i_1}\zeta^{(1)}_{i_2}\ldots\,\zeta^{(m-1)}_{i_{m}},\eeq
which is the standard volume form $\sigma_0,$ on $S\subset\BR^{m*}$ - cf. \cite[p. 165]{greubI}. The formula
(\ref{eq:rn}) has been proven for $V$ in Ref. \cite[Proposition 6]{jad06c}. The proof for $V^*$ goes much the
same way. \qed\vskip10pt
\subsection{The representations $R_r$ of $G$ on $C^\infty(S)$}
In this subsection we will define a family of (infinite dimensional) representations $R_r,\, r\in\BR,$ of the group $G=SO_0(1,m)$ on the space
of (smooth) functions on $S.$ The representations $R_r$ will be closely related to the representations induced
from representations of a little group, the subgroup of $G$ that stabilizes the point $\bn=(0,0,\ldots\,0,1)\in
S,$ except that our representations will be, at this stage, non-unitary (cf. however Proposition \ref{prop:uni} below), so that
we will skip the part of the induced
representation theory (Radon-Nikodym derivative) that is usually added there to guarantee unitarity. We will
adapt the definition of the induced representation as given, for instance, in Ref. \cite[Ch. 5, p. 174, Eq. (36),
and p. 215, Theorem 6.7]{var}\footnote{We will also skip measure--theoretical considerations, as we work in the
category of smooth functions and actions.}

\begin{defn} Let $W$ be a finite--dimensional vector space and let
$\gamma$ be a cocycle. Given $p\in\{0,\dots ,m-1\},$ the following formula defines the representation $R^p_\gamma$ of $G$ on
the space $\bigwedge^p(S;W)$ of $W$--valued $p$--forms on $S$  \beq R^p_\gamma(\Lambda) \psi ={\varrho_{\Lambda^{-1}}}^*(\gamma(\Lambda,\cdot)\psi
).\label{eq:rl}\eeq The representation $R^p_\gamma: \Lambda\mapsto R^p_\gamma(\Lambda)$ is called the
representation determined by the cocycle $\gamma.$ When $\gamma=\gamma_r,$ as in Eq. (\ref{eq:gammar}), then
$R^p_{\gamma_r}$ will be denoted simply by $R^p_r.$\\  When $W=\BR$ we will use the brief notation
$\bigwedge^p(S)$ for $\bigwedge^p (S;\BR).$ Notice that $\bigwedge^p(S;W)=\bigwedge^p(S)\otimes W.$
\label{def:rep}
\end{defn}
\subsection{The spaces $\mathcal{Y}_r^p$ of equivariant forms}
We will denote by $\mathcal{Y}_r^p(S;W)$ the space of $R^p_r$--equivariant maps from $P$ to $\bigwedge^p(S;W).$
Explicitly, if $\psi: P\ni e\mapsto \psi[e]\in \bigwedge^p(S;W)$ is such a map, then \beq\psi[e\Lambda](\bn)=
\gamma_r(\Lambda,\bn )^{-1}\,(\varrho_{\Lambda}^*\psi[e])(\bn),\quad \Lambda\in G,\,e\in P,\, \bn\in
S.\label{eq:equivf}\eeq We will simply write $\mathcal{Y}_r^p$ for $\mathcal{Y}_r^p(S;\BR).$ The next Proposition follows immediately from the definitions and from Eq.
(\ref{eq:gadd}).
\begin{prop}
If $\phi\in\mathcal{Y}^p_r$ and $\psi\in \mathcal{Y}^q_s(S;W)$ then $(\phi\wedge\psi )\in
\mathcal{Y}^{p+q}_{r+s}(S;W).$\qed \label{prop:ymul}\end{prop} For the proof of the next Theorem we will need the following Lemma.
\begin{lem} The following are examples of elements of spaces $\mathcal{Y}_r^p$:
\begin{itemize}
\item[\rm{(i)}] The map $\hat{\phi},$ $P\ni e\mapsto\hf[e]={\hat \phi}_e\in C^\infty(S;V^*),$ defined in Eq. (\ref{eq:phie}) -- cf. also Diagram \ref{eq:diag} --
is in $\mathcal{Y}^0_{-1}(S;V^*).$
\item[\rm{(ii)}] Given a vector $v\in V$ and a frame $e\in P$ consider the function $f_v[e]:S\rightarrow \BR$ defined
by $f_v[e](\bn)=v[e]^\alpha \omega_\alpha=v[e]^0+v[e]^i \omega_i.$ Then $f: v\mapsto f_v$ is a linear map from $V$ to
$\mathcal{Y}^0_{-1}.$
\item[\rm{(iii)}] The constant map $\sigma: e\mapsto \sigma[e]=\sigma_0$ defined in Lemma \ref{lem:sigma} and assigning to each
$e\in P$ the standard $SO(m)$--invariant volume form $\sigma_0$ on $S,$ is a member of
$\mathcal{Y}^{(m-1)}_{(1-m)}.$
\end{itemize}
\label{lem:exmpl}\end{lem}
\pf
(i) The statement follows by a straightforward calculation. Let $e=(e_\alpha)$ be a basis in $P,$ and let
$(e^\alpha )$ be the dual basis. For $\Lambda\in G$ we have
\begin{eqnarray*} \hf[e\Lambda](\bn)=\omega_\alpha (e\Lambda)^\alpha &=&(e\Lambda)^\alpha \omega_\alpha=\aij{\Lambda^{-1}}{\alpha}{\beta}\omega_\alpha e^\beta \\
&=& (\aij{\Lambda^{-1}}{0}{0}+\aij{\Lambda^{-1}}{i}{0}\omega_i)\left(e^0+\frac{\aij{\Lambda^{-1}}{0}{j}+\aij{\Lambda^{-1}}{i}{j} \omega_i}{\aij{\Lambda^{-1}}{0}{0}+\aij{\Lambda^{-1}}{i}{0} \omega_i}\,e^j\right)\\
&=& \gamma_1(\Lambda,\bn)\hf[e](\varrho_\Lambda(\bn))=\gamma_1(\Lambda,\bn)(\rho_\Lambda^*\hf[e])(\bn).
\end{eqnarray*}
(ii) Notice that $f_v[e](\bn) = \langle e^\alpha,v\rangle \omega_\alpha=\langle e^\alpha
\omega_\alpha,v\rangle=\langle \hf[e](\bn),v\rangle,$ therefore the results follows
from (i).\\
(iii) We first notice that by Lemma \ref{lem:gadd} we have $(\gamma_r)^{-1}=\gamma_{-r}.$ The statement follows
then from Eqs. (\ref{eq:rn}) and (\ref{eq:equivf}). \qed\vskip10pt
\begin{thm}
Let $\phi,\psi$ be in $\mathcal{Y}_{m/2}^0.$ Then, for every $v\in V,$ the $(m-1)$--form $\phi\wedge\psi\wedge
f_v\wedge\sigma$ is in $\mathcal{Y}_{0}^{(m-1)}$ and the following integral $I(\phi,\psi,v)$ does not depend on
the frame $e\in P$: \beq I(\phi,\psi,v)=\int_S \phi[e]\wedge\psi[e]\wedge f_v[e]\wedge\sigma[e].\eeq The map
$I(\phi,\psi): v\mapsto I(\phi,\psi,v)\in\BR$ is linear in $v$, and defines a bilinear form on
$\mathcal{Y}_{m/2}^0$ with values in $V^*.$
\end{thm}
\pf
It follows from Proposition \ref{prop:ymul} and from Lemma \ref{lem:exmpl} (ii),(ii)  that $\phi\wedge\psi\wedge
f_v\wedge\sigma$ is in $\mathcal{Y}_{0}^{(m-1)},$ therefore $\phi[e]\wedge\psi[e]\wedge f_v[e]\wedge\sigma[e]$ is
an $(m-1)$--form on $S$ that is independent of the choice of $e\in P.$ The rest of the theorem is na immediate
consequence of the definitions. \qed\vskip10pt
\subsection{The spaces  $\mathcal{Y}^0_r.$ as spaces of homogeneous functions on $C^{*+}$}
While working with the spaces $\mathcal{Y}_r^p$ of equivariant form--valued functions is sufficient for technical
purposes, it is convenient to have a geometrical interpretation of the results. For this end we will only need a
geometrical interpretation of the spaces $\mathcal{Y}^0_r.$

In Section \ref{sec:gqef} above we introduced the spaces of quantities of type $R,$ where $R$ is a representation
of the structure group $G$ on a vector space $F.$ In our case we take $G=SO(1,m),$ and we will identify, in this
section, the spaces $\mathcal{Y}^0_r$ with the spaces $Y_{-r}(C^{+*})$ of homogeneous functions of degree $-r$ on
$C^{+*}.$
\begin{defn}
For each $r\in \BR$ let $Y_r(C^{+*})$ be the vector space of smooth real functions on $C^{+*},$ homogeneous of
degree $r.$ That is, a smooth function $f:C^{+*}\rightarrow\BR$ is in $Y_r(C^{+*})$ if and only if $f(\lambda
p)=\lambda^r f(p)$ for all $\lambda\in\BR^+,\, p\in C^{+*}.$ In particular, for every $v\in V$ the function
$f_v:$ \[C^{+*}\ni p\mapsto f_v(p)=<v,p>=v^\alpha p_\alpha\in \BR \] is homogeneous of degree $1.$
\end{defn}
\begin{prop}\label{prop:yr}
The function space $Y_r(C^{+*})$ is naturally isomorphic to the space $\mathcal{Y}^0_{-r}.$ More precisely, if
$f\in Y_r(C^{+*}),$ then $\hf^*(f)$ defined by \beq \hf^*(f)[e](\bn)= f(\hf[e](\bn))\eeq is in
$\mathcal{Y}^0_{-r}.$
\end{prop}
\pf
Indeed, with the notation as above, we have
\begin{eqnarray*}
\hf^*(f)[e\Lambda](\bn)&=&f(\hf[e\Lambda](\bn))=f(\gamma_1(\Lambda,\bn)(\varrho_\Lambda^*\hf[e])(\bn))\\
&=&\gamma_1(\Lambda,\bn)^r f((\varrho_\Lambda^*\hf[e])(\bn))\\
&=&\gamma_r(\Lambda,\bn)\varrho_\Lambda^*(\hf^*(f)[e])(\bn).
\end{eqnarray*}
The result follows then from Eq. (\ref{eq:equivf}). \qed\vskip10pt
\section{The Kairon field}
Let $M$ be an $(m+1)$--dimensional manifold. Let ${\mathcal V}$ be a vector bundle over $M,$ with a typical fiber
$\BR^{m+1},$ ${\mathcal V}=\bigcup_{x\in M} V_x,$ endowed with a $G=SO_0(1,m)$ structure. Its dual vector bundle will
be denoted by $\mathcal{V}^*.$ We will denote by ${\mathcal F}=\bigcup_{x\in M} F_x$ the bundle of linear frames
of ${\mathcal V},$ and by ${\mathcal P}=\bigcup_{x\in M} P_x$ the principal sub--bundle of ${\mathcal F}$ that
defines the $G$ structure on ${\mathcal V}.$ We will denote by $\mathcal{C}^{+*}$ the bundle of positive isotropic
cones in the fibres of $\mathcal{V}^*.$ Let $\Theta$ be a $1$--form over $M$ with values in ${\mathcal V},$ thus,
for each $\xi\in T_xM,$ $\Theta_x$ is a linear map $\Theta_x: T_xM\rightarrow V_x.$ We will call $\Theta$ the
soldering form. \begin{rem} In a simplified version of the theory one can identify ${\mathcal V}$ with the
tangent bundle of $M.$ In this simplified case $\Theta$ would be the identity map. However, we are proposing a
more general formulation, which allows us to treat gravity as a composite field, along the lines developed in
Ref. \cite{jad82}, where we have discussed ``gauge theories of gravity", so that the soldering form can as well
be not of a maximal rank over certain parts of $M.$\end{rem}

We will assume that ${\mathcal P}$ is equipped with a principal connection. The corresponding exterior covariant
derivative\footnote{For a good, modern introduction to differential geometrical concepts cf. the recent monograph by
 Mari\'{a}n Fecko \cite{fecko}}, acting on differential forms with values in the associated bundles will be denoted as $D.$ Of
particular interest for us will be the vector bundle $\cav^*,$ which can be thought of as being associated to ${\mathcal P}$ via the representation $R'$ of $G$ on
$E^{(1,m)*}$ given by: \beq R'(\Lambda): E^{(1,m)}\ni (p_\mu)\mapsto \aij{(\Lambda^{-1})}{\nu}{\mu}\,p_\nu\quad
\Lambda\in G,\eeq and its sub--bundle ${\mathcal C}^{+*}$ of positive isotropic cones : ${\mathcal
C}^{+*}=\bigcup_{x\in M} C_x^{+*}\subset {\mathcal V}_x^*.$
\subsection{Recall of differential geometric concepts}\label{sec:rdgc}
Let $(P,\pi,M,G)$ be a principal bundle with base manifold $M$ and structure group $G.$ Let $R$ be a left action
of $G$ on a manifold $Q.$ We denote by $P\times_R Q$ the associated bundle (with a typical fiber $Q$), and by
$\Gamma(P\times_R Q)$ the space of its (smooth) sections. We denote by $C^\infty(P,Q)^R$ the space of all smooth
mappings $f: P\rightarrow Q$ that are $R$--equivariant, that is such that $f(pg)=R(g^{-1})f(p)$ holds for $p\in
P$ and $g\in G.$ As in Section \ref{sec:gqef} there is a natural one--to--one correspondence between the elements
of $\Gamma(P\times_R Q)$ and those of $C^\infty(P,Q)^R.$ More generally, let $R$ be a representation of $G$ on a
vector space $W,$ and let $\Omega(M,P\times_R W)$ be the graded algebra of $P\times_R W$ valued differential
forms on $M.$ Let $\Omega_{hor}(P,W)^G$ be the graded algebra of $W$-valued horizontal, $G$--equivariant,
differential forms on $M.$ Then there is a canonical isomorphism $q^\sharp : \Omega^k(M,P\times_R W)\rightarrow
\Omega^k_{hor}(P,W)^G.$ For every $\Phi\in \Omega^k(M,P\times_R W),$ $p\in P,$ $\zeta_1,\ldots,\zeta_k\in T_p P$
we have: \beq p\cdot (q^\sharp (\Phi )_p(\zeta_1,\,\ldots,\zeta_k))= \Phi_{\pi
(p)}(d\pi_p(\zeta_1),\ldots,d\pi_p(\zeta_k)).\eeq For $k=0$ the isomorphism $q^\sharp$ reduces to the isomorphism
between $\Gamma(P\times_R Q)$ and $C^\infty(P,Q)^R.$ For details see e.g. Ref \cite[Sec.
(21.12),(22.14)]{michorf}.

Let $\mathfrak{g}$ denote the Lie algebra of $G$ (carrying the adjoint representation of $G$) and let $\omega$ be a principal
connection on $P.$ In particular we have that $\omega\in \Omega^1(P,\mathfrak{g})^G.$ We will denote by $D_\omega$ (or simply by $D,$ when it is clear from the context which
principal connection is being used) the exterior covariant derivative $D: \Omega^k(P,W)\rightarrow
\Omega^{k+1}(P,W).$ By abuse of notation we will denote by the same symbol $D$ the exterior derivative acting on
forms with values in associated bundles, that is on elements of $\Omega(M,P\times_R W).$ For details see e.g.
Ref. \cite[Sec. (22.15)]{michorf}. We recall the following result, adapted from Ref. \cite[Proposition VIII, p.
254]{greubII}:
\begin{prop} Let $\phi: W_1\times \ldots \times W_l\rightarrow W$ be an $l$--linear map and let $\Phi_i\in
\Omega^{k_i}_{hor}(P,W_i)^G$  be $W_i$--valued horizontal, $G$--equivariant differential forms of degree
$k_i,\,(i=1,\ldots ,l).$ Then \beq D[\phi_*(\Phi_1,\ldots
,\Phi_l)]=\sum_{i=1}^l(-1)^{k_1+\ldots+k_{i-1}}\phi_*(\Phi_1,\ldots ,D\Phi_i,\ldots ,\Phi_l).\eeq
\qed\label{prop:mapphi}\end{prop}
\begin{rem} For $\Phi_i$ of the form $\Phi_i=\Psi_i\otimes w_i,$ $\Psi_i\in \Omega^{k_i}_{hor}(P,\BR)^G,$ $w_i\in W_i,$
the mapping $\phi_*$ is defined as \beq \phi_*(\Phi_1,\ldots,\Phi_l)=(\Psi_1\wedge\ldots \wedge
\Psi_l)\otimes\phi(w_1,\ldots,w_l).\eeq It extends by linearity for a general case. In other words $\phi_*$ is the map $\phi$ applied to the values
of the forms, not to their arguments.
%Equivalently:
%\beq \phi_*(\Phi_1,\Phi_2,\ldots ,\Phi_l)[e]=\phi(\Phi_1[e]\wedge\Phi_2[e]\wedge\ldots\wedge\Phi_l[e]),\quad e\in P.\eeq
\end{rem}

%\begin{cor} Let $R_i,R$ be representations of $G$ on vector spaces $W_i,W,$ and let $\tilde{\phi}:
%(P\times_{R_1} W_1)\times_M\ldots,\times_M (P\times_{R_l} W_l)\rightarrow (P\times_{R} W)$ be an $l$--linear map
%of vector bundles over $M.$ Let ${\tilde\Phi}_i\in \Omega^{k_i}(M,P\times_{R_i}W_i),$ then \beq
%D[{\tilde\phi}_*({\tilde\Phi}_1,\ldots
%,{\tilde\Phi}_l)]=\sum_{i=1}^l(-1)^{k_1+\ldots+k_{i-1}}{\tilde\phi}_*({\tilde\Phi}_1,\ldots
%,D{\tilde\Phi}_i,\ldots ,{\tilde\Phi}_l).\eeq
%\end{cor}
%\pf
%It is enough to use the fact that $q^\sharp({\tilde \Phi}_i)=\Phi_i$ and then Ref. \cite[Sec.
%(22.15)]{michorf}.\qed\vskip10pt
\subsection{The Kairon bundle $\Upsilon[{\mathcal P}].$}
In this paper we will discuss only real fields of spin $0.$ The case of spin $\frac{1}{2}$ will be discussed in a forthcoming paper.\\
Let $Y$ be the space of functions on the positive isotropic cone $C_0^{+*}\subset
E^{(1,m)*},$ homogeneous of degree $-\frac{m}{2},$ \footnote{The
reason for choosing this particular degree of homogeneouity will be evident from Proposition \ref{prop:fe}.} and let $T$ be the natural representation of $G$ on $Y$: \beq (T(\Lambda) f)(p)= f(p\Lambda).\eeq We
denote by $\Upsilon$ the associated vector bundle ${\mathcal P}\times_T Y$ and by $\Gamma(\Upsilon)$  the $C^{\infty}(M)$ module  of local
sections of $\Upsilon.$ Even if the fibres of $\Upsilon$ are infinite dimensional function spaces, we will apply
the standard constructions of differential geometry as they can be easily generalized and applied without changes
to this particular case - cf. e.g. \cite[Ch. 5 and references therein]{michorgt}.

The following proposition follows immediately from our previous discussion:\footnote{When necessary the standard precautions
concerning local rather than global operations should be applied.}
\begin{prop}A section
$\psi\in\Gamma(\Upsilon)$ can be interpreted in five different ways, namely as:
\begin{enumerate}
\item[{\rm (a)}] An equivariant function on $P$ with values in $Y,$ $\Psi[e\Lambda]=T(\Lambda^{-1})\Psi[e].$
\item[{\rm (b)}] A function ${\tilde \psi}$ on $M$ with values in $\Upsilon,$ ${\tilde \psi} (x)\in \Upsilon_x.$
\item[{\rm (c)}] A real--valued function ${\hat \psi}$ on $\mathcal{C}^{+*}$ homogeneous of degree $-\frac{m}{2}.$
\item[{\rm (d)}] A real valued function ${\tilde\Psi}:P\times S\rightarrow \BR,$ that is equivariant in the following sense:
\beq {\tilde\Psi}(e\Lambda,\bn)= \left(
\aij{\Lambda^{-1}}{0}{0}+\aij{\Lambda^{-1}}{i}{0}\omega_i\right)^{-\frac{m}{2}}{\tilde\Psi}(e,\varrho_\Lambda(\bn)).\label{eq:equivar1}\eeq
\item[{\rm (e)}] An equivariant function ${\hat\Psi}: e\mapsto {\hat\Psi}[e]$ on $P$ with values in $C^\infty(S)$ defined
as ${\hat\Psi}[e](\bn)={\tilde\Psi}(e,\bn).$
\end{enumerate}
\qed\label{prop:ways}\end{prop}
\subsection{Generalized torsion}
Our connection is in the bundle $P,$ not in the bundle of frames of $M.$ We define (generalized) torsion $T$ as
$T=D\Theta,$ the exterior derivative of the soldering form. Thus $D\Theta$ is a two--form on $M$ with values in
the vector bundle $\mathcal{V}.$ In local coordinates, it has components $T_{\mu\nu}^a,$ and there is no way of
``lowering" the index $a,$ unless the soldering form $\Theta$ is bijective. In those regions of $M$ where
$\Theta$ is bijective we can use it to define a metric tensor and a unique affine connection on $M$ by demanding
that $\Theta$ is parallel with respect to the pair of connections: the connection in the bundle ${\mathcal P}$ and the
connection in the frame bundle of $M.$ \footnote{For more details concerning this last point cf. Ref.
\cite{jad82}}
\subsection{The Field equations}
Usually in physical theories the field equations are being deduced from a variational principle. Then
conservation laws follow either from field equations or from invariance principles. In our case we will postulate
the field equations directly, because it is not clear at this time whether some kind of a variational principle
leading to these field equations can be construed. The conservation law will follow directly from the field
equations.

Let us first discuss a particular consequence of Proposition \ref{prop:mapphi} by taking $W_1=W_2=C^\infty(S),$
$W_3=\BR^{m+1}$ and the map $\phi: W_1\times W_2\times W_3\rightarrow \BR$ given by: \beq \phi
(\psi_1,\psi_2,w)=\int_S \psi_1(\bn)\psi_2(\bn)w^\alpha \omega_\alpha \sigma_0 (\bn).\eeq We take $\Psi_1,\Psi_2\in
\Gamma (\Upsilon).$ Then ${\hat\Psi_1},{\hat\Psi_2}$ are $0$--forms on $P$ with values in $W_1=W_2=C^\infty(S).$
The soldering form $\Theta$ can be considered as an equivariant horizontal form ${\hat\Theta}=q^\sharp (\Theta)$
on $P$ with values in $\BR^{m+1}.$
\begin{prop}
With $\Psi_1,\Psi_2\in \Gamma(\Upsilon ),$  $x\in M,$ $\xi\in T_xM$ and $e\in P_x$  let ${\hat
j}_{\Psi_1,\Psi_2}[e](\xi)$ be defined by the formula \beq {\hat j}_{\Psi_1,\Psi_2}[e](\xi)=\int_S
{\hat\Psi_1}[e](\bn){\hat\Psi_2}[e](\bn)\Theta(\xi)^\alpha \omega_\alpha \sigma_0(\bn).\label{eq:j12}\eeq Then ${\hat
j}_{\Psi_1,\Psi_2}[e]$ is independent of $e$ and defines a $1$--form $j_{\Psi_1,\Psi_2}$ on $M.$ Moreover, if we assume that $\Psi_1,\Psi_2$
satisfy the field equations: \beq \Theta\wedge D\Psi_i =\frac{1}{2} \Psi_i D\Theta,\, (i=1,2),\label{eq:kairons}\eeq
then the form $j_{\Psi_1,\Psi_2}$ is closed: \beq d\,j_{\Psi_1,\Psi_2}=0.\label{eq:closed}\eeq
\label{prop:fe}\end{prop}
\pf It is clear from the definitions that ${\hat j}_{\Psi_1,\Psi_2}=\phi_*({\hat \Psi}_1,{\hat \Psi}_2,{\hat \Theta}).$
Since ${\hat j}_{\Psi_1,\Psi_2}$ is $\BR$--valued we have that $d{\hat j}_{\Psi_1,\Psi_2}=D{\hat j}_{\Psi_1,\Psi_2}.$ On the other hand, from
Proposition \ref{prop:mapphi} we have that \beq D{\hat j}_{\Psi_1,\Psi_2}=\phi_*(D{\hat \Psi}_1,{\hat\Psi}_2,{\hat \Theta})+\phi_*({\hat \Psi}_1,D{\hat\Psi}_2,{\hat \Theta})+\phi_*({\hat \Psi}_1,{\hat\Psi}_2,D{\hat \Theta}).\label{eq:dphi}\eeq
Now, using the field equations (\ref{eq:kairons}), and skipping the argument $[e],$ we have that
\beq \begin{array}{rcl}\phi_*(D{\hat\Psi}_1,{\hat\Psi}_2,{\hat\Theta})&=&\int_S \left(D{\hat\Psi}_1\wedge{\hat\Psi}_2\wedge{\hat\Theta}^\alpha\right)(\bn)\omega_\alpha\sigma_0(\bn)\\
&=&-\int_S \left({\hat\Psi}_2\wedge{\hat\Theta}^\alpha\wedge D{\hat\Psi}_1\right)(\bn)\omega_\alpha\sigma_0(\bn)\\
&=&-\frac{1}{2}\int_S \left({\hat\Psi}_1\wedge{\hat\Psi}_2\wedge D{\hat\Theta}^\alpha\right)(\bn)\omega_\alpha \sigma_0(\bn).\\
\end{array}\eeq
Similarly \beq \phi_*({\hat\Psi}_1,D{\hat\Psi}_2,{\hat\Theta})=-\frac{1}{2}\int_S \left({\hat\Psi}_1\wedge{\hat\Psi}_2\wedge D{\hat\Theta}^\alpha\right)(\bn)\omega_\alpha \sigma_0(\bn),\eeq
and finally
\beq \phi_*({\hat\Psi}_1,{\hat\Psi}_2,D{\hat\Theta})=\int_S \left({\hat\Psi}_1\wedge{\hat\Psi}_2\wedge D{\hat\Theta}^\alpha\right)(\bn)\omega_\alpha \sigma_0(\bn).\eeq Thus, taking into account Eq. (\ref{eq:dphi}) we have (\ref{eq:closed}).\qed\vskip10pt
\begin{rem}Introducing a $U(1)$ principal bundle with a principal connection and replacing the exterior covariant derivative $D$ by the exterior
covariant derivative including the $U(1)$ connection, it is easy to generalize our field equations and get current conservation $dj=0$ for charged kairons.\end{rem}
\section{The case of a flat Minkowski space}
In this section we will study the solutions of the field equations (\ref{eq:kairons}) for the case where $M$ is
the flat space $E^{(1,m)}$ endowed with  the natural orientation and time--orientation, and with the natural zero connection (thus zero torsion). For the bundle $P$
we take the bundle of oriented and time--oriented orthonormal frames, therefore $\Theta$ will be the identity map. If $e\in P$ then $<e_\mu,e_\nu>=\eta_{\mu\nu},$ where $\eta=\mathrm{diag }(-1,+1,\ldots ,+1).$ We will endow $M$ with the
standard coordinate systems (we will call them Lorentz frames) related one to another by transformations from the
proper inhomogeneous Lorentz group, the semi--direct product of the proper ortochronous Lorentz group $SO_0(1,m)$ and
the group of translations $\BR^{(m+1)}.$ We will use the standard terminology of special relativity: light
cone, time--like and space--like vectors etc.

The field equations (\ref{eq:kairons}), in a Lorentz frame $x^\mu$ take the form \beq (\omega_\mu \partial_\nu
-\omega_\nu\partial_\mu)\Psi(x,\bn)=0,\label{eq:kflat}\eeq where $\omega_0=1,\bn^2=1.$ It follows that for each $\bn,$ and
for each bivector $f=(f^{\mu\nu}),$ the solutions are constant on the trajectories of the vector field
$f^{\mu\nu}\omega_\mu\partial_\nu.$ These vector fields, for different $f$ commute and they span the $m$--dimensional
hyperplane $X_\bn$ that is annihilated by the $1$--form $\omega.$ Since $\omega$ is a light--like co--vector, the plane
$X_\bn$ is isotropic: it is generated by the light--like vector $\omega^\mu=\eta^{\mu\nu}\omega_\nu$ and $m-1$ space--like
vectors. We will first show that every solution of Eqs. (\ref{eq:kflat}) is uniquely determined by the initial
data $g(t,\bn)$ on the time axis $x^0=t,\bx=0,\, t\in\BR.$
\begin{prop}{(Initial value problem)}
Let $g(t,\bn)$ be an arbitrary function on $\BR\times S^{m-1}.$ There exists a unique solution $\Psi(x,\bn)$ of
the field equations (\ref{eq:kflat}) that coincides with $g$ on the time axis $\bx=0.$
\end{prop}
\pf
Fix an isotropic co--vector $\omega=(1,\bn).$ Let $(y^0,\by)$ be an arbitrary point in $M$ that is not on the time
axis (that is, $\by\neq 0).$ The line connecting a point $(x^0,0)$ on the time axis to this point is given by
$x^0(s)=sy^0+(1-s)x^0,\, \bx(s)=s\by.$ The tangent vector to this line has components $(y^0-x^0,\by),$ and the
value of $\omega$ on the tangent vector is then $y^0-x^0+\bn\cdot\by.$ It takes the value zero if and only if
$x^0=y^0+\bn\cdot\by.$ It follows that \beq\Psi(y^0,\by)=\Psi(y^0+\bn\cdot\by,{\bf 0}).\label{eq:propag}\eeq Therefore $\Psi$ is
determined by its values on the time axis. The values of $\Psi$ on the time axis can be arbitrary, because any
two different points on the axis are connected by a time--like vector, while all vectors on an isotropic plane
are either space or light--like.
\vskip10pt
It is easy to generalize the above property to a more general class of time--like paths.
\begin{cor}
Let $\mathcal{T}$ be the set of all time--like paths $\gamma(s),\, s\in \BR$ with the property that
$\gamma$ has a non--empty intersection with each maximal isotropic plane. Then, for each $\gamma\in \mathscr{T},$
every solution $\Psi$ of the
field equations (\ref{eq:kflat}) is uniquely determined by its values on $\gamma.$
\end{cor}
\pf
The proof goes as in the proposition above by first noticing that any two points on $\gamma$ are connected by a time--like
interval, while no two points on a maximal isotropic plane are connected by such an interval.\qed
\vskip10pt
\begin{figure}[h!]
\begin{center}
    \leavevmode
      \includegraphics[width=6cm, keepaspectratio=true]{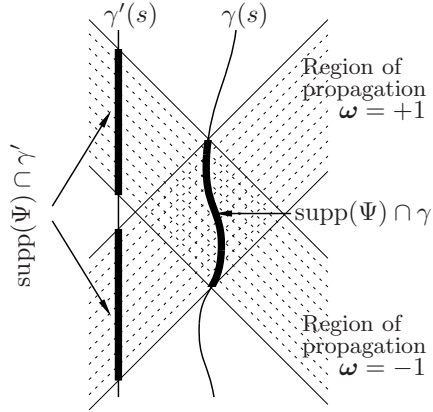}
\end{center}
  \caption{Kaironic Field propagation in $1+1$ dimensions.}
\end{figure}
\begin{rem}
A typical example of a time--like trajectory that is not in $\mathscr{T}$ is a world--line of a uniformly accelerated
observer (hyperbolic motion), such as obtained from the trajectory $\gamma(s)=(s,\mathbf{0})$ by a special conformal transformation.
\end{rem}
\subsection{The Hilbert space of solutions}
Let $\gamma$ be a path of class $\mathscr{T}$ and suppose that $\Psi$ is a solution of the field equations with the property
that it vanishes outside of a compact part of $\gamma.$ That means there exists $s_0>0$ such that $\Psi(\gamma(s),\bn)=0$
for $|s|>s_0.$ Then, owing to the propagation formula (\ref{eq:propag}), $\Psi$ vanishes in the interior of the past
last cone with apex at $(-s_0,\gamma(-s_0))$ and in the interior of the future light cone with apex at $(s_0,\gamma(s_0)).$
If $\gamma'$ is another path in $\mathscr{T},$ then $\gamma'$ has also the property that $\Psi$ vanishes on $\gamma'$ outside of
a compact set.
\begin{prop} Let $\mathscr{D}(\Upsilon)$ be the vector space of solutions $\Psi$ of the field equations (\ref{eq:kflat}) with the property
that $\mathrm{supp}(\Psi)\cap \gamma$ is compact for every path $\gamma\in\mathscr{T}.$ Then, for any $\Psi_1,\Psi_2\in\mathscr{D}(\Upsilon)$ the integral:
\beq \langle\Psi_1,\Psi_2\rangle = \int_{S^{m-1}}d\sigma_0(\bn)\int_{-\infty}^{+\infty}ds\,{\dot\gamma}(s)^\alpha\omega_\alpha\, \Psi_1(\gamma(s),\bn)\Psi_2(\gamma(s),\bn)\label{eq:scprod}\eeq
does not depend on the choice of the path $\gamma\in\mathscr{T}$ and defines a real pre--Hilbert structure in $\mathscr{D}(\Upsilon).$
\end{prop}
\pf
The integral (\ref{eq:scprod}) is nothing but the integral of the one--form $j$ given by Eq. (\ref{eq:j12}) over the path $\gamma .$
If $\gamma'$ is another path in $\mathscr{T},$ then it is always possible to make a closed oriented loop by adding segments $l$ and $l'$ that are outside
of $\mbox{supp}(\Psi)$ (cf. Fig. 1). Then the integral of $j_{\Psi_1,\Psi_2}$ over this loop vanishes owing to the fact
$j_{\Psi_1,\Psi_2}$ is a closed one--form (cf. Eq. (\ref{eq:closed}). It follows the integrals of $j_{\Psi_1,\Psi_2}$ over
$\gamma$ and over $\gamma'$ are equal. By choosing $\gamma(s)=(s,\bf{0})$ we have ${\dot\gamma(s)}^\alpha\omega_\alpha=1,$
therefore the scalar product is positive definite and thus it defines a pre--Hilbert structure on $\mathscr{D}(\Upsilon).$
\qed
\vskip10pt
\subsection{Poincar\'{e} invariance}
Our presentation will be here more sketchy than in the previous sections and we will use the shortcuts, the notation
and the rigor typical for the papers on theoretical physics.

The Poincar\'{e} group $SO_0(m,1)\circledS \BR^{m+1}$ acts on the bundle $P$ of orthonormal frames of the Minkowski
space by bundle automorphims that preserve the flat connection. Therefore it acts on the space of solutions of
the field equations (\ref{eq:kflat}) preserving the invariant scalar product (\ref{eq:scprod}). It is instructive to
see this action explicitly and to identify the infinitesimal generators of this action in terms of the Hilbert
space $L^2(\BR,S^{m-1})$ of initial data on the $x^0$ axis of a fixed Lorentz reference frame.

%While the invariance under translations is evident, the analysis of the action of Lorentz transformations
%requires some care.
%
%Let $\phi_\Lambda:M\rightarrow M$ be a Lorentz transformation:
%\beq \phi_\Lambda(x)^\mu =\aij{\Lambda}{\mu}{\nu}x^\nu,\, \Lambda\in SO_0(m,1).\eeq
%Denoting by $\partial=(\partial_\mu)$ the orthonormal frame tangent to the coordinate lines $x^\mu$, the transformation
%$\phi$ maps this frame into another orthonormal frame $\tilde{\partial},$ tangent to the transformed coordinate lines, \beq\tilde{\partial}_\mu=(d\phi_\Lambda)(\partial_\mu)=\aij{\Lambda}{\nu}{\mu}\partial_\nu=(\partial\Lambda)_\mu.\eeq
%Now, the group $\mbox{Aut}\,(P)$ of all automorphisms of the principal bundle $P$ acts on vector valued functions on $P$ by linear
%transformations. If $\phi$ is in $\mbox{Aut}\,(P),$ and if $f$ is a $W$--valued function on $P,$ then
%$(U_\phi f)(e)=f(\phi^{-1}(p)).$ Applying this formula to $d\Phi_\Lambda$ and a function $\Psi$ in $\mathscr{D}(\Upsilon),$
%we obtain:
%\beq (U_\Lambda \Psi)(x,\partial)=\Psi(\Lambda^{-1}x,\partial\Lambda^{-1}).\eeq
%If we introduce $\Psi(x,\bn)$ defined by $\Psi(x,\bn)=\Psi(x,\partial)(\bn)$ then, making use of the equivariance property
%(\ref{eq:equivar1}) we obtain:
%\beq (U_\Lambda\Psi)(x,\bn)=\left(
%\aij{\Lambda}{0}{0}+\aij{\Lambda}{i}{0}\omega_i\right)^{-\frac{m}{2}}\Psi(\Lambda^{-1}x,\varrho_\Lambda^{-1}\bn).\eeq
The simplest way to obtain the explicit expressions for the group action is by using the formulation (c) of Proposition \ref{prop:ways}.
For simplicity we will use the symbol $\Psi(x;\omega)$ (resp. $\Psi(x,\bn)$) to denote
$\hat{\Psi}$ (resp. the restriction of $\hat{\Psi}$ to the section $\omega_0=1$ of the positive light--cone $\mathcal{C}^*$).
\subsubsection{Lorentz invariance}
Let $\phi_\Lambda:M\rightarrow M$ be a Lorentz transformation:
\beq \phi_\Lambda(x)^\mu =\aij{\Lambda}{\mu}{\nu}\,x^\nu,\, \Lambda\in SO_0(m,1).\eeq This transformation
induces the action on $\mathcal{C}^{+*},$ which we will denote by the same symbol, $\phi_\Lambda(\omega)_\mu=\omega_\nu\aij{\Lambda^{-1}}{\nu}{\mu}.$
If $\Psi(x;\omega)$ is a solution of the field equations (\ref{eq:kflat}) then the transformed solution
$U_\Lambda\Psi$
is given by:
\beq (U_\Lambda \Psi)(x^\mu;\omega_\alpha)=(\phi_{\Lambda^{-1}*}\Psi)(x^\mu;\omega_\alpha)=\Psi(\aij{\Lambda^{-1}}{\mu}{\nu}\,x^\nu;(\omega\Lambda)_\alpha).\eeq
The next step is to set $\omega=(1,\bn)$ and to write
\beq \omega_\beta\aij{\Lambda}{\beta}{\alpha}=\omega_\alpha\aij{\Lambda}{\alpha}{0}\frac{\omega_\beta\aij{\Lambda}{\beta}{\alpha}}{\omega_\alpha\aij{\Lambda}{\alpha}{0}}.\eeq
The fraction term is on the section of the light--cone by the plane $\omega_0=1$ and its space part is, taking into account
Eq. (\ref{eq:rholambda}), equal to $\varrho_{\Lambda^{-1}}(\bn).$ On the other hand, as a function of
$\omega,$ the function $\Psi$ is homogeneous of degree $-\frac{m}{2}.$ Therefore we have that
\beq (U_\Lambda \Psi)(x^\mu;\bn)=((\omega\Lambda)_0)^{-m/2}\,\Psi(\aij{\Lambda^{-1}}{\mu}{\nu}\,x^\nu;\varrho_{\Lambda^{-1}}(\bn)).\eeq
Now we restrict $(U_\Lambda \Psi)(x^\mu,\omega_\alpha)$ to the $x^0$ axis $x^i=0.$ We have then $\aij{\Lambda^{-1}}{0}{\nu}\,x^\nu=\aij{\Lambda^{-1}}{0}{0}\,x^0,$
and $y^i\stackrel{def}{=}\aij{\Lambda^{-1}}{i}{\nu}\,x^\nu=\aij{\Lambda^{-1}}{i}{0}\,x^0,$ therefore
\beq (U_\Lambda\Psi)(x^0,\mathbf{0};\bn)=((\omega\Lambda)_0)^{-m/2}\,\Psi\left((\aij{\Lambda^{-1}}{0}{0}x^0,\by;\varrho_{\Lambda^-1}(\bn)\right)\eeq
Since $\Psi$ is a solution of the field equations, we can use the propagation formula (\ref{eq:propag}) to obtain the following
formula on the $x^0$ axis:
\beq (U_\Lambda\Psi)(x^0;\bn)=((\omega\Lambda)_0)^{-m/2}\Psi (\aij{\Lambda^{-1}}{0}{0}x^0+\varrho_{\Lambda^{-1}}(\bn)\cdot\by;\varrho_{\Lambda^{-1}}(\bn))\eeq
We now compute the term
\beq A=\varrho_{\Lambda^{-1}}(\bn)\cdot\by=\frac{(\omega\Lambda)_i}{(\omega\Lambda)_0}\aij{\Lambda^{-1}}{i}{0}x^0.\eeq
Notice that we have:
\begin{eqnarray*}\omega_\alpha\aij{\Lambda}{\alpha}{i}\aij{\Lambda^{-1}}{i}{0}&=&\omega_\alpha(\aij{\Lambda}{\alpha}{\beta}\aij{\Lambda^{-1}}{\beta}{0}-\aij{\Lambda}{\alpha}{0}\aij{\Lambda^{-1}}{0}{0})\\
&=&\omega_\alpha(\delta^\alpha_0-\aij{\Lambda}{\alpha}{0}\aij{\Lambda^{-1}}{0}{0})\\
&=&1-(\omega\Lambda)_0\aij{\Lambda^{-1}}{0}{0}.
\end{eqnarray*}
Therefore \[A=\frac{x^0}{(\omega\Lambda)_0}-\aij{\Lambda^{-1}}{0}{0}x^0,\]
which gives us the final expression for the Lorentz transformed solution in terms of the initial data:
\beq (U_\Lambda\Psi)(x^0;\bn)=((\omega\Lambda)_0)^{-m/2}\,\Psi \left(\frac{x^0}{(\omega\Lambda)_0};\varrho_{\Lambda^{-1}}(\bn)\right).\label{eq:ulambda}\eeq
While the invariance of the scalar product (\ref{eq:scprod}) follows from our geometrical considerations, because of
rather non--standard nature of the transformations, it is instructive to verify the unitarity of $U_\Lambda$
directly.
\begin{prop}Transformations $U_\Lambda$ given by Eq. (\ref{eq:ulambda}) preserve the scalar product
(\ref{eq:scprod}) on $\BR\times S^{m-1}$:
\beq\langle \Psi_1,\Psi_2\rangle=\int_{S^{m-1}}\sigma_0(\bn)\int_{-\infty}^{\infty}dx^0\, \Psi_1(x^0;\bn)\Psi_2(x^0;\bn).\eeq
\label{prop:uni}\end{prop}
\pf
It is sufficient to show that transformations $U_\Lambda$ preserve the norm. We have
\[ \Vert U_\Lambda\Psi\Vert^2=\int_{S^{m-1}}\sigma_0(\bn)\int_{-\infty}^{\infty}dx^0\,(\omega\Lambda)_0^{-m}\Psi^2 \left(\frac{x^0}{(\omega\Lambda)_0};\varrho_{\Lambda^{-1}}(\bn)\right).\]
Introducing a new variable $y^0=\frac{x^0}{(\omega\Lambda)_0},\, dx^0=(\omega\Lambda)_0\,dy_0,$ we obtain
\[ \Vert U_\Lambda\Psi\Vert^2=\int_{S^{m-1}}\sigma_0(\bn)\int_{-\infty}^{\infty}dy^0\,(\omega\Lambda)_0^{-m+1}\Psi^2 \left(y^0;\varrho_{\Lambda^{-1}}(\bn)\right).\]
We can now introduce a new variable ${\bn}'=\varrho_{\Lambda^{-1}}(\bn).$ Taking into account the fact
that owing to the Eq. (\ref{eq:rn}) we have  $$(\omega\Lambda)_0^{-m+1}\sigma_0(\bn)=\gamma(\Lambda^{-1},\bn)_{1-m}\sigma_0(\bn)=\sigma_0({\bn}'),$$ which completes the
demonstration.\qed
\vskip10pt
\subsubsection{Translation invariance}
The action of time translations is evidently unitary, therefore we need to consider only space translations,
With $\ba\in\BR^m$ we have
\beq (U_\ba\Psi)(x^0;\bn)=\Psi(x^0,-\ba;\bn)=\Psi(x^0-\ba\cdot\bn;\bn).\eeq
The unitarity follows from translation invariance of the Lebesgue measure $dx^0.$
%\subsubsection{Infinitesimal generators}
\section*{Acknowledgments}
Thanks are due to Pierre Angl\`{e}s and Mari\'{a}n Fecko for reading the manuscript and for pointing out several misprints.


\begin{thebibliography}{99}
\bibitem{barbour} Barbour,~Julian,: {\em The End of Time\,}, Oxford University Press, Oxford, 2000.
\bibitem{prigogine} Prigogine,~Ilya: {\em From Being to Becoming\,}, W.H. Freeman and Company, New York, 1980
\bibitem{blaja96} Blanchard,~Ph.,  Jadczyk,~A.: {\em Relativistic Quantum Events\,}, Found. Phys. 26, 1669-1681 (1996)
\bibitem{ceja77} Ceg{\l}a,~W., and Jadczyk,~A.: {\em Causal Logic of the Minkowski Space\,}, Commun. Math. Phys. 57, 213--217 (1977); Borowiec,~A., and Jadczyk,~A.: {\em Covariant Representations of the Causal Logic\,}, Lett. Math. Phys. 3, 255--257 (1979)
\bibitem{rodvaz} Rodrigues,~Jr,,~W.~A. and Vaz,~Jr., J.: {\em Subluminal and superluminal solutions in vacuum of the Maxwell equations and the massless Dirac equation\,}, in {\em Proceedings of the International Conference on
The Theory of the Electron\,}, Eds. Jaime Keller and Zbiniew Oziewicz, Adv. Appl. Clifford Algebras, Proc. Suppl. 7, (S1), 1997.
\bibitem{sternberg} Sternberg,~S.: {\em Lectures on Differential geometry\,}, Prentice Hall, 1964
\bibitem{schouten} J.~A.~Schouten, {\em Tensor Analysis for
Physicists\,}, Dover, NY, 1954
\bibitem{greubI} Greub,~W., Halperin, S.,  and Vanstone, B.: {\sl Connections, Curvature, and Cohomology}, Vol. I., Academic Press, New York, 1972, Exercise 4, p. 273
\bibitem{jad06c}Jadczyk,~A.: {\em Quantum Fractals on n-spheres.
Clifford algebra approach\,}, {{\protect \tt http://arxiv.org/abs/quant-ph/0608117}}
\bibitem{schwartz} L.~Schwartz, {\em  Analyse II -- Calcul Diff\'erentiel et \'Equations Diff\'erentielles,\,}
(Hermann, Paris 1992)
\bibitem{var} Varadarajan,~V.~S.: {\em Geometry of Quantum Theory\,}, Second Edition, (Springer, New York 1985)
%\bibitem{br} Barut,~A.~O., Raczka,~R.: {\sl theory of Group Representations and Applications\,}, Second revised edition, World
%Scientific , Singapore 1986
\bibitem{jad82} Jadczyk,~A.: {\em Vanishing Vierbein}, {{\protect \tt http://arxiv.org/abs/gr-qc/9909060}}
\bibitem{fecko} Fecko,~M.: {\em Differential Geometry and Lie Groups for Physicists,\,} (Cambridge University Press, Cambridge 2006)
\bibitem{michorf} Michor,~P.~W.: {\em Topics in Differential Geometry\,}, Draft from December 28, 2006, {{\protect \tt http://www.mat.univie.ac.at/~michor/listpubl.html}}
\bibitem{greubII} Greub, W., Halperin, S.,  and Vanstone, B.:
{\sl Connections, Curvature, and Cohomology}, Vol. II., Academic Press, New York, 1973
\bibitem{michorgt} Michor,~P.: {\em Gauge Theory for Fiber Bundles\,}, vol. 19 of Monographs and Textbooks in Physical Science. Lecture Notes, Bibliopolis, Naples, 1991
\end{thebibliography}
\end{document}